\documentclass[12pt,preprint]{aastex}

\lefthead{Escala}
\righthead{M$_{BH}-{\sigma} \rm$ cutoff}
\def\msun{M_{\odot}}

\usepackage{lineno}
\begin{document}

\title{On the Fate of Little Red Dots }

\author{Andr\'es Escala$^1$, Lucas Zimmermann$^2$, Sebasti\'an Valdebenito$^1$, Marcelo C. Vergara$^3$, Dominik R.G. Schleicher$^4$, Mat\'ias Liempi$^4$}
\affil{1 Departamento de Astronom\'{\i}a, Universidad de Chile, Casilla 36-D, Santiago, Chile.}
\affil{2 Department of Astronomy, Yale University, New Haven, CT 06511, USA}
\affil{3 Astronomisches Rechen-Institut, Zentrum fur Astronomie, University of Heidelberg, Monchhofstrasse 12-14, 69120, Heidelberg, Germany}
\affil{4 Dipartimento di Fisica, Sapienza Universit\`a di Roma, Piazzale Aldo Moro 5, 00185 Rome, Italy}
\affil{Correspondence to: aescala@das.uchile.cl}

\begin{abstract}
We study the %show /discuss  that 
%nuclear stellar clusters avoid the collisional-dominated regime, while massive dark objects are well into stellar collision are dynamically relevant.
%the stability of the inner region of galaxies mainly depends on the velocity dispersion $\sigma \rm$, with a maximum limit around $\rm \sigma_c \sim 400  \,  km s^{-1}$, in which   galactic nuclei becomes unstable (for higher velocity dispersions than $\rm \sigma_c$, thru runaways stellar collisions) with possible fate being the formation of a massive black hole. 
stability and possible fates of Little Red Dots,  under the stellar-only interpretation of their observational features. This is performed by a combination of %studying 
analyzing the relevant timescales  in their  stellar dynamics and also, the application of recent  numerical results on the evolution of the densest stellar systems. We find that %some of
 these objects   typically  have $t_{age} \sim \, t_{coll} < t_{relax}$, therefore, in an unexplored  regime  never  observed before  for a stellar system and potentially, highly unstable to runaway collisions. We study different scenarios for the evolution of Little Red Dots and conclude  that in a fair fraction of those systems, the formation of a massive black hole by runaway stellar collisions seems unavoidable, in all the possibilities studied within the  stellar-only interpretation. 
 This evolutionary path   would naturally explain many of the problematic characteristics of Little Red Dots, including that these objects are probably transient in the  history of the Universe, that most of them would not emit X-rays since they would not yet have become massive black holes, and once they do, they would  constitute a significant portion of the mass of the Little Red Dots. 
We conclude that Little Red Dots are the most favourable  known places to find a recently formed massive black hole seed, or   in the process of formation, most probably formed directly within the supermassive range. %  and the most appealing places to host the massive seeds.
\end{abstract}


\section{Introduction}

The \textit{James Webb} Space Telescope (JWST) revolutionized  our vision of the early universe, detecting z $>$ 14 galaxies (e.g., R. Naidu et al. 2025), and uncovering a new population of high-redshift sources that had eluded earlier telescopes: the Little Red Dots (LRDs; D. D. Kocevski et al. 2023; Y. Harikane et al. 2023; J. Matthee et al. 2024), which are characterized to be extremely compact and red. More striking,  detections of LRDs concentrate only in the redshift range 4 $<$ z $<$ 8 (%being  at those redshifts roughly 100 times more abundant than UV-selected quasars; 
Akins et al. 2024; D. D. Kocevski et al. 2025), making this population of galaxies observable  during a cosmic period of only $\lesssim$ 1 Gyr, i.e. less than  one orbital period for a Milky Way-type galaxy (comparable in stellar mass to the most massive LRDs), thus appearing as almost transitory objects  even for galactic timescales  and certainly, for cosmological ones. LRDs population is  roughly 100 times more abundant than UV-selected quasars at z $>$ 4 (Akins et al. 2024; D. D. Kocevski et al. 2025), but individual LRDs and their analogs have only been discovered in small numbers at z $\leq$ 4, being their number density %al lower redshift still
 a matter of intense research  (Ma et al 2025; Zhuang et al 2025).

Two main interpretations  have been suggested for LRDs, each peculiar and not previously observed before JWST, resembling  the  puzzle raised in the 1960s  by the discovery of the quasars. The first one, that LRDs host a super-Massive Black Hole (MBH)  at their centers  due to the presence of broad Balmer emission lines (R. Maiolino et al. 2024; J. E. Greene et al. 2024). However,  most  LRDs are undetected in X-rays, even in deep stacking analyses (T. T. Ananna et al. 2024; F. Pacucci \& R. Narayan 2024;  P. Madau \& F. Haardt 2024; R. Maiolino et al. 2025). Assuming that the black holes still are present, they seem to be over-massive respect  to their host  (F. Pacucci et al. 2023; K. Inayoshi \& K. Ichikawa 2024; E. Durodola et al. 2025), compared to the masses  expected  from extrapolating the local relations (J. Magorrian et al. 1998, L. Ferrarese \& D. Merritt 2000; K. Gebhardt et al. 2000). 

The second interpretation is that they are intensely star-forming dusty galaxies (Akins et al 2024), but the cores reach extreme stellar densities being the densest stellar systems ever observed (Guia et al. 2024). The latter 
comes from  their compactness, one of the most  striking features of the LRDs,  with typical effective radii  $\sim$100 pc are (on average) an order of magnitude more compact  than the smallest galaxies previously observed. Moreover, the  widths of
the Balmer broad lines can also be explained by the stellar velocity dispersion at the cores of these system (A. Loeb
2024; J. F. W. Baggen et al. 2024),  generating velocities of $\sim$1500 km $s^{-1}$, around  factor 4 larger than the maximum velocity dispersion documented in the nearby Universe ($\rm \sigma = 444 \, km s^{-1}$; Salviander et al 2008), whose long term stability is  probably not viable (A. Escala 2026).

The goal of this paper is to show that regardless of the interpretation, the expected fate of LRDs  is that they will eventually host a MBH (or become one in the most extreme cases). This is because (even) under the  stellar only interpretation, LRDs  extreme stellar densities implies that a fair fraction (up-to  unity)  of their inner regions will unavoidably  end in a MBH, expected  also to be over-massive  with respect to the host. This is based in the scenario outlined in A. Escala (2021), in which MBHs are formed in-situ in galactic nuclei as (partially or totally) failed stellar systems. This paper is organized as follows: we start studying the stability of LRDs under the stellar-only interpretation in \S2, analyzing  the relevant timescales  for stability in  \S2.1 and possible unstable outcomes in  \S2.2. In   \S3, we will discuss a summary of the results and conclusions achieved.

\section{Stability of LRDs in the Stellar-Only Interpretation%: relevant timescales}
}

In this section we will study the  long term stability of LRDs,  in the context of  a scenario for MBH seeds that forms by direct collapse in galactic nuclei from a failed stellar system, that is unstable due to runaway stellar collisions. This was  proposed  in Escala (2021),  who studied the stability of stellar systems in galactic nuclei and compared with the observed properties of galactic nuclei (MBHs and Nuclear Stellar Clusters; NSCs), being able to explain the relative trends observed in the masses, efficiencies, and scaling relations between MBHs %massive black holes and nuclear stellar clusters. 
and NSCs. Several numerical results have latter supported the viability of  this scenario, focused on   different aspects of it  (M.C. Vergara et al 2023, 2024, 2025a,b), including the first state-of-art  (and one particle per star) N-body simulation able to form a Very Massive Star (VMS; which  subsequently collapses to a  MBH seed) that reaches the $\rm \sim 10^5\, \msun$ mass scale (M.C. Vergara et al  2025a) from  runaway stellar collisions, following up on the suite of DRAGON simulations (L. Wang et al 2016, M. Arca-Sedda et al 2024). %in particular, being able to show that the efficiency in the formation of the MBH by runaway stellar collisions dramatically increases, for systems with average collision timescales  $\rm t_{coll}$ that are comparable to the age of the system   $\rm t_{age$. 

\subsection{Relevant Timescales in the Stability of Stellar Systems}

There are two distinct  regimes when the stellar systems in galactic nuclei may  become globally unstable, being determined by relative values of the relaxation $\rm t_{relax}$  and collision $\rm t_{coll} $ timescales  (Escala 2021): $\rm t_{relax} < t_{coll}$ and $ \rm t_{coll} <  t_{relax}$, when these (average) timescales are dynamically relevant (i.e. comparable or shorter than the age of the system $\rm t_{age}$). Stellar systems with $\rm t_{relax} < t_{coll} \,\, (<  t_{age})$ are initially globally unstable,  but are able to readjust  before collisions are triggered overall in the system, since the cluster will expand before  (on  few relaxation times $\rm%\sim
 t_{relax}$; Henon 1965; Gieles et al. 2012), due to two-body encounters which guarantees the global stability of the stellar system. However, on the same timescale the core will undergo collapse (Cohn 1980; Portegies Zwart \& McMillan 2002) enhancing the runaway  collisions within the core. These runaway collisions are expected to lead the formation of a BH within the core %(REF)
  and the predicted final state is the coexistence of a MBH and a NSC  (Escala 2021). 
  
This first regime, $\rm t_{relax} < t_{coll} $%\,\, (<  t_{age})$
, has been recently tested numerically by means of direct N-body simulations (M.C. Vergara et al 2023, 2024, 2025a,b),  in particular, being able to show that the efficiency in the formation of the MBH by runaway stellar collisions exponentially increases (before saturation towards 1), with MBH formation efficiencies up to 50\% of the final cluster mass, for systems with average collision timescales  $\rm t_{coll}$ that are comparable to the age of the system   $\rm t_{age}$ (M.C. Vergara et al 2023). This result was also  shown to be valid (M.C. Vergara et al 2024)  for a diversity of  stellar systems (NSCs, globular cluster (GCs), and ultra-compact dwarf galaxies (UCDs))   and methods (both numerical simulations and observations fullfil it, encompassing various initial conditions, initial mass functions, and evolutionary scenarios). In principle, with   MBH efficiencies $\sim$  10$-$50\% %in the proposed mechanism, 
a $\rm 10^{6-7} \msun$ stellar cluster could form a MBH seed directly in the super-massive regime,  compared to alternative mechanisms  that are also based on runaway stellar collisions and the collapse of a VMS, but predict MBHs seeds in the intermediate-mass regime (i.e. A. Askar et al 2021, E. Gonz\'alez-Prieto et al 2024, A. Rantala et al. 2025).

In  another regime  when  $\rm t_{coll} <  t_{relax}$% \,\, (<  t_{age})$
, the stellar system will not able to adjust before collisions are triggered everywhere, which are highly inelastic and efficiently  convert the  orbital motions that support the cluster (against its self-gravity) into heat (that subsequently can be  radiated away), leading to a possible collapse of the system as a whole on to a MBH. In this regime, no stellar system should survive in the long term and for that reason, we will subsequently call  it `Forbidden Stellar Zone', particularly  to the  region of cluster parameters (i.e. mass and radius), where no stellar system is expected to survive. Unfortunately, simulations under these conditions  are  extremely  expensive numerically, so restrictive that no direct N-body simulations has been performed yet for systems in  this regime, but simulations M.C. Vergara et al (2023, 2024, 2025a,b) are suggestive since even in the $\rm t_{relax} < t_{coll}$  regime,  an  exponential increase is found that  reaches BH  efficiencies upto 50\% of the final  (MBH + stellar) system.

One of the advantages of the simple picture outlined in  Escala (2021), is that it is possible to evaluate the previously mentioned  regimes for observed systems with  measured and  radius R and masses M, since the  relevant timescales can be expressed in just those two quantities,  under few assumptions such as being  virialized systems (i.e. with a velocity dispersion $ \sigma = \sqrt{GM/R}\,$). The collision timescale  for  virialized systems can be expressed as $t_{coll} = \sqrt{\frac{R}{GM(n\Sigma_0)^2}\,}$, where $R$ and $M$ are the radius and the mass of the system, $G$ is the gravitational constant, $\Sigma_0$ is the effective cross-section and $n$ is the number density of stars (= $\eta M/M_*R^3 $, with $\eta = 3/4\pi $ for a uniform spherical system). The
effective cross section  $\Sigma_0$, counting  the gravitational focusing is $\Sigma_0 = 16 \sqrt{\pi} R_*^2(1+ \Theta)$, where $\Theta=9.54 (\frac{M_* R_\odot}{M_\odot R_*})(\rm\frac{100\, kms^{-1}}{\sigma})^2$ is the Safronov number, with $M_*$ and $R_*$ being the typical masses and radius of the stars, respectively, and $\sigma$ is the (rms) stellar velocity dispersion, which again can be expressed in terms of M and R as $ \sigma = \sqrt{GM/R}$ using the virial theorem. Therefore,  $t_{coll} $ can be expressed only in terms of M and R for a system composed of stars with typical masses $M_*$ and radius $R_*$. Same case for the relaxation time $t_{relax}= \frac{0.1N}{\ln{N}} t_{cross}$, where $N$ is the total number of stars, which is simply $N = M/M_*$  if the system is composed of equal-mass stars,  and $t_{cross}=\sqrt{R^{3}/GM}$ is the crossing time for a virialized system, thus  $t_{relax}$  can also be expressed only  in terms of M and R, for a system composed of stars with typical masses $M_*$.

\begin{figure}[h!]
\begin{center}
\includegraphics[width=11.9cm]{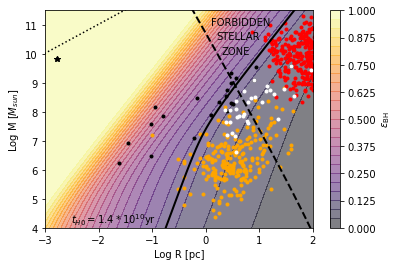}
\caption{Total masses M and sizes R in color points for the following systems:  resolved MBHs (black), unresolved MBHs (white), NSCs (orange) and LRDs (red). The solid black curve follows the condition $\rm t_{coll} (M, R) = t_{H0}$ for a system of a given mass M and radius R (at the current age of the universe  $\rm t_{H0}  = 1.4 \times 10^{10}$ yrs) and the dashed one,  the condition $\rm t_{relax} (M, R) = t_{H0}$. The color diagram denotes the black hole efficiency expected from $\epsilon_{BH}  = \left(1 + \exp\left[-4.63 \left( \log\left(M/M_{crit}\right) - 4 \right) \right]\right)^{-0.1}$ (M.C. Vergara et al 2023, 2024, 2025b), with $\rm M_{crit}$ coming from the condition $\rm t_{coll} \,(M_{crit}) = t_{H0}$. Those efficiencies match the expectations for the MBH formation scenario outlined in Escala (2021), where MBHs are formed from (partially/totally) failed stellar systems, by approaching towards 1 for resolved MBHs (black points) and being  NSCs mainly in the range between 0.05 and 0.15   (coexistence of a MBH and a NSC; Escala 2021). The region called `Forbidden Stellar Zone'  fulfill the instability defined by $\rm t_{coll}  <  t_{H0} \, (< t_{relax})$, where no stellar system is expected to survive, being most LRDs at the boundary of such instability (within one order of magnitude if we consider their upper limits in total effective radius).}
\label{F1}
\end{center}
\end{figure}

Figure \ref{F1} shows both  the collision and relaxation timescales,  in a mass-radius  plane compared to the current age of the Universe (i.e. $\rm t_{age} = t_{H0} = 1.4 \times 10^{10} yr$, the maximum relevant time as an upper limit  for the real ages) and for a stellar system (mainly) composed of solar mass stars. The solid black line displays the condition $t_{coll} =   t_{H0}$, therefore, systems on the  left side of the solid black line  fulfill the condition  $t_{coll}  <  t_{H0}$, thus collisions may be triggered everywhere in the system, contrarily to systems on the right side of the black line ($t_{coll}  >  t_{H0}$), where collisions should not be dynamically relevant (outside the core, if this collapses) and the system can be long term stable.  The case is similar for  the dashed black one, that denotes  the condition $t_{relax} =   t_{H0}$, dividing the regions  $t_{relax}  <  t_{H0}$ (left side to the dashed line) and  $t_{relax}  >  t_{H0}$ (right side). The orange points displays the total masses and effective radius in NSC (Georgiev et al 2016). Black points denotes resolved MBHs and the white  ones, unresolved MBHs (Gultekin et al. 2009). The black dotted line (upper left)  denotes Schwarzschild radius for a given mass, where real sizes of  black holes should be, like the  Event Horizon Telescope (2019) detection for M87 (black star just below the dotted line).

The solid black line  in Figure \ref{F1} ($t_{coll} =   t_{H0}$), clearly %divides 
separates the population of resolved MBHs  (black points) with the NSCs (orange) and unresolved MBHs (white ones, with properties diluted to the ones of the background  stellar system and thus representative of it instead of the MBH; Escala 2021). The simplest interpretation is that stellar systems born on the left side of the solid black line, do not survive since it become globally unstable to runaway collisions and collapse to a MBH, in agreement  with resolved MBHs  (black points) being in that region of   Figure \ref{F1} and contrarily, survived stellar systems  born on the right  side of the solid black line, in agreement with the location of orange and white points (NSCs and background  stellar system of  unresolved MBHs) in  Figure \ref{F1}. However,  the discussed points are mainly below the intersection of solid and dashed lines, thus eventhough systems that born with 
 $t_{coll}  <  t_{H0}$ are initially globally unstable, they have even shorter relaxation times  ($\rm t_{relax} < t_{coll}$) 
and  the stellar system  will be able to expand and survive before  collapsing as a whole onto a MBH. Only their cores may become unstable and  lead to the formation of a BH by  runaway collisions within the core, being 
 the predicted final state for those systems  the possible coexistence of a MBH and a NSC (Escala 2021).

%MISMA HISTORIA  2021 with a `Forbidden Stellar Zone' 

%As mentioned before, t
This simple picture was  further supported recently by  several direct N-body numerical calculations. M.C. Vergara et al (2023) showed that  the efficiency in the formation of the MBH by runaway stellar collisions exponentially  increases as  $\rm t_{coll}$ becomes comparable to  $\rm t_{age}$, independent of having (even) shorter relaxation times, mainly because relaxation processes produces core collapse (and envelope expansion) but the efficiency is controlled by the collision time. M.C. Vergara et al (2024)  expanded this finding to different stellar systems, including not only NSCs, but also GCs and UCDs, covering different initial conditions, stellar initial mass functions, and evolutionary paths, which can be fitted with an asymmetric sigmoidal curve, given by  $\epsilon_{BH}  = \left(1 + \exp\left[-4.63 \left( \log\left(M/M_{crit}\right) - 4 \right) \right]\right)^{-0.1}$
%.  $\epsilon_{BH}  = \left(1 + \exp\left[-k \left( \log\left(M/M_{crit}\right) - x_0 \right) \right]\right)^{a}$, where $k = 4.63$ controls the steepness of the transition, $x_0 = 4$ sets its location, and $a = -0.1$ determines the smoothness of the function 
(M.C. Vergara et al 2025b), where $M_{crit}$ is the critical mass for a system with collision time comparable  to the age of the system  (i.e. $\rm t_{coll} \,(M_{crit}) = t_{H0}$, for the purposes of  this analysis). Figure \ref{F1} also displays  such black hole formation efficiency $\epsilon_{BH} $ in the (background) color gradient, being   NSCs mainly in the range between 0.05 and 0.15, approaching towards 1 for resolved MBHs (i.e. if they were stellar clusters, should rapidly collapse to a MBH driven by runaway stellar collisions), in an overall agreement with their observed properties and also, supporting  the scenario outlined in Escala (2021), where MBHs are formed from (partially/totally) failed stellar systems.

The position of LRDs in Figure \ref{F1}, denoted by the red points (data taken from Akins et al 2024), is different from any other stellar system previously observed at lower z, since LRDs appear at the boundary of the region we called `Forbidden Stellar Zone', located  above  the intersection of solid and dashed black lines, in the left side of the solid black line where no stellar system is expected to survive. LRDs are typically only one order of magnitude %(the accuracy of an order of magnitude estimate) 
 from the solid black line (if we consider as real their upper limits in total effective radius, probably closer in reality), compared with typical galaxies of similar mass (e.g $~10^{11} \msun$ in $>$ 10kpc) that are located several  orders of magnitude to the right of the solid black line in Figure 1, which is %probably 
the main reason why the collision timescale have been traditionally ignored in galactic dynamics.  Such region fulfills the instability defined by $t_{coll}  <  t_{H0}$,  left side of the solid black line like the densest NSCs borned, but with  a quantitative difference: in the mass range of LRDs (masses $ \gtrsim 10^9 \msun$),  the condition $\rm  t_{coll} < t_{relax} $ it is  typically fulfilled furthermore. Therefore, LDRs with $t_{coll}  <  t_{H0}$ are on a different regime from unstable NSCs,  without being able to expand due to relaxation processes, since $t_{H0} < t_{relax}$ for LRDs, thus candidates to be  in the unexplored regime of %globally 
collapsing as a whole due to the  loss of orbital support  by runaway  stellar collisions,   expected when $t_{H0} \sim \, t_{coll} < t_{relax}$.

%Moreover
In addition, errors in mass estimates are in the order of 0.3 dex (under the stellar only interpretation; Akins et al 2024), making little difference on the proximity of LRDs to the `Forbidden Stellar Zone'. Only if a MBH is  assumed already at the center for most LRDs, their stellar mass estimates decreases to values lower than $~10^{10} \msun$ (Jones et al. 2025), but since  their total effective radius are upper limits in most cases, lower real radius %will 
may bring  LRDs back to the proximity to the `Forbidden Stellar Zone',  something expected unless other effects such surface brightness dimming operates (Pacucci \& Loeb 2025). Even in the latter 
%pessimistic
 scenario, since all self-gravitating systems are centrally concentrated, the  instability discussed  here should  still be %even
 more violent  in the core than in the whole system. In the next section we will study  possible outcomes for LRDs as a whole system and within their cores.

\subsection{%Possible  Fates: 
Globally Unstable  Outcomes}

In this section we will study the  stability and possible fates for LRDs, exploring three possible avenues for such extremely dense stellar systems, with the aim of studying the viability for MBH formation  triggered by runaway stellar collisions,   under the stellar-only interpretation for LRDs observational features.

We will make an estimate for the possible MBH masses expected for LRDs from the numerical results of M.C. Vergara et al (2023, 2024, 2025a, b), since they evolved  the densest stellar systems simulated so far, thus the closest to the inferred densities  in LRDs under the stellar-only interpretation, potentially shedding  light on the evolution and fates of those systems. Nevertheless, they are not strictly applicable  to these systems, mainly because for most  LDRs $t_{coll} <  t_{relax}$ (i.e. located above the intersection of dashed and solid lines in Fig 1), aversely to the  numerical results of M.C. Vergara et al (2023, 2024, 2025a, b) that are strictly valid  only in the regime $ t_{relax} < t_{coll}$ and also,  can only considered upper limits for BH masses expected in  such regime since  primordial binaries are not included in those numerical studies. This fact has quantitative differences,  since when  $  t_{coll} <  t_{relax}$, the system will not able to adjust (due to two-body relaxation) before collisions are triggered everywhere, leading to a possible collapse of the system as a whole on to a MBH (if  in addition, $ t_{coll} <  t_{H0}$ is fulfilled, which is not the case for most LRDs if the measured upper limits in radius correspond to their real sizes). On the other hand, no core collapse is expected for LRDs, since  $ t_{H0} < t_{relax}$  for theses systems and therefore, the initial growth of the MBH should  slow down. Therefore, to apply M.C. Vergara et al (2023, 2024, 2025a, b) results to LRDs, it is implicitly assumed that both processes compensates.  This assumption could also be supported by the non-dependence yet reported  of the BH  formation efficiency $\epsilon_{BH}$  on  $\rm  t_{relax}$ (i.e. negligible or secondary compared to the dependence on $\rm  t_{coll}$ found in M.C. Vergara et al 2023, 2024), being  $\epsilon_{BH}$ also determined  for a diversity of stellar systems, not only the one simulated    in M.C. Vergara et al (2023, 2025a, b), including information from  observed systems that might form their BH in a  regime more similar to the  one in LRDs.

Red points in Fig 2a shows the LRDs in a mass radius diagram. The solid black line displays again the condition $ t_{coll} =   t_{H0}$  and the dashed one,  denotes the condition $t_{relax} =   t_{H0}$, both computed at the age of the Universe  $ t_{H0} = 0.6\times 10^9$yr, which corresponds to the time of appearance of LRDs in the universe (z $\sim$ 8). The position of LRDs  avoids the  forbidden stellar zone, with only one exception, besides all  LRDs fulfilled the condition  $ t_{coll} <  t_{relax}$ (i.e. masses larger than the intersection of the solid and dashed black curves). Figure 2a  also displays the predicted  black hole masses, in a  contour plot  with color gradient in the background,  using the formation efficiency $\epsilon_{BH}$  computed numerically in M.C. Vergara et al (2025b), namely $M_{BH}  = \epsilon_{BH} \times M = M \times \left(1 + \exp\left[-4.63 \left( \log\left(M/M_{crit}\right) - 4 \right) \right]\right)^{-0.1}$.  The bulk of the predicted masses for MBHs from  the $\epsilon_{BH}$  in M.C. Vergara et al (2025b), are in the range between $\sim 5\times  10^6\msun$ and $\sim 10^{10}\msun$, incidentally   in the same range observed for MBHs in the local universe (Krolik 1999). Also,  this formation efficiency in M.C. Vergara et al (2025b) predicts  MBHs that will  be over-massive respect to their host compared to the masses in the local relations, ranging from few percents of their host mass, to  a fraction more similar to the one expected in the interpretation that LRDs already host a MBH ($\epsilon_{BH} \sim$ 0.1; F. Pacucci et al. 2023; K. Inayoshi \& K. Ichikawa 2024; E. Durodola et al. 2025), for LRDs  approaching   towards the solid black line in Fig 2a (i.e. $\epsilon_{BH} (M/M_{crit}=1)$= 0.157).

%a BH efficiency $\epsilon_{BH}$= 0.01-0.1 for $M/M_{crit} \sim 0.01$, leading to masses in the supermassive range ($10^6-10^9 \msun$) OR a contour plot with efficiencies/bh masses in   Fig 1b?. 
%King Profile? $R^{1/4}$? u otro...

%\begin{figure}[h!]
\begin{figure}
%\begin{center}
%\begin{subfigure}%{\textwidth}%
  \centering
 \includegraphics[width=0.51 \linewidth]{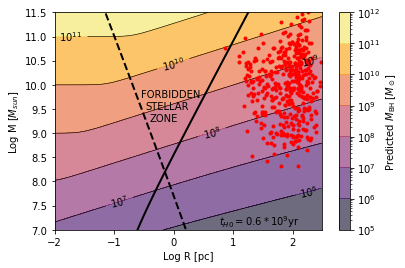}
 %\caption{Lifespan mutation burden plotted against versus the  body mass. End-of-life mutation burden equals to  3000 is denoted  by the  dashed black line.}
 %\label{fig:sfig1}
%\end{subfigure}

%\begin{subfigure}%{\textwidth}
 % \centering
 \includegraphics[width=0.4301  \linewidth]{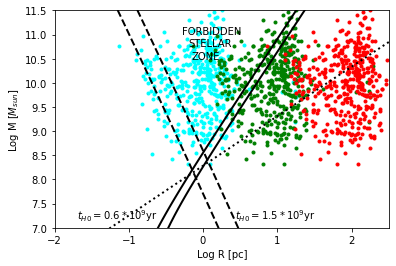}
% \caption{Lifespan mutation burden plotted against versus the  body mass. End-of-life mutation burden equals to  3000 is denoted  by the  dashed black line.}
%\label{fig:sfig2}
%\end{subfigure}

%\begin{subfigure}%{\textwidth}
%  \centering
 \includegraphics[width=0.4301  \linewidth]{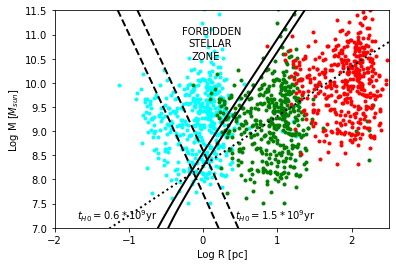}
%\caption{Top: Middle:Bottom:}
% \label{fig:sfig3}
%\end{subfigure}

\caption{(a)  Total masses M and radius R in red points for LRDs,  the solid black curve following the condition $\rm t_{coll} (M, R) = t_{H0}$ for the  age of the universe at z$\sim$8 ($ t_{H0}  = 0.6 \times 10^{9}$ yrs)  and the dashed one,  the condition $\rm t_{relax} (M, R) = t_{H0}$. The color diagram denotes the predicted black hole mass  $\rm M_{BH}$  expected from $\rm M \times \left(1 + \exp\left[-4.63 \left( \log\left(M/M_{crit}\right) - 4 \right) \right]\right)^{-0.1}$ (Vergara et al 2023, 2024, 2025b), %with $\rm M_{crit}$
predicting $\rm M_{BH}$ an approximate range between $\rm  5\times  10^6\msun$ and $10^{10}\msun$. 
(b) The positions of LRDs for the measured  upper limit radius in red, in green LRDs positions assuming that their real effective radii corresponds to the 10\% of the upper limits and in cyan, assuming that corresponds to the 1\% of the  upper limits in radius. The solid  and dashed black curves denotes  the same condition as in (a), but  at the cosmic age  of appearance     ($ t_{H0} = 0.6\times 10^9$yr)  and disappearance ($ t_{H0} = 1.5\times 10^9$yr) of LRDs.  In the case of the green points, a fraction enters to the  `Forbidden Stellar Zone' ($ t_{coll} <  t_{H0} < t_{relax}$), while the cyan points it  is  more extreme, with most LRDs are in such situation.  (c) Same as (b), now assuming that it is the core radius of the LRDs the one equals to the 10\% (green) or 1\% (cyan) of their measured effective radii, with most of the cores  entering to the  `Forbidden Stellar Zone' for the case where LRDs core radius equals to the 1\% of their measured  radii.}
\label{F2}
%\end{center}
\end{figure}

In addition to the fact that the M.C. Vergara et al (2023, 2024, 2025a, b) results are not strictly applicable to LRDs, there are two  additional  reasons for exploring  a different interpretation for the evolution and possible fates of LRDs. The first one,  is their transitory nature on a cosmological context, since this population observed during a cosmic period of only  a Gyr in total (i.e. the timespan between  z $\sim$ 4 and %$\sim$ 
8), possibly having individual lifespan of $\sim 10^8$ yrs (or  less). The second one,  is a direct consequence of their compactness: besides LRDs have typical effective radii  $\sim$100 pc, in most cases these are only upper limits, possibly entering to the `Forbidden Stellar Zone' if their real  effective radii are considerable smaller than the measured upper limits.  In Fig 2b, we  start exploring the consequences  of a lower  effective radii than the measured  upper limits and for that, we computed the conditions of   $t_{coll} =   t_{H0}$ (solid black lines) and  $t_{relax} =   t_{H0}$ (dashed black lines), at the cosmic age  of appearance     ($ t_{H0} = 0.6\times 10^9$yr or z $\sim$ 8)  and disappearance ($ t_{H0} = 1.5\times 10^9$yr or z $\sim$ 4) of LRDs in the universe (D. D. Kocevski et al. 2024). The dotted line corresponds to the locus of the condition $t_{coll} =  t_{relax} $ (for solar mass stars) at different cosmic times $  t_{H0}$, which can be computed   analytically   and equivalent  to the condition  $\rm R/R_{\star} = \sqrt {\frac{1.2 (1+ \Theta )}{\sqrt \pi lnN}}\, N$, being N=$M/M_{\star}$ and noting that  $\Theta \approx$ 0.5 when  $t_{coll} $ approaches $ t_{relax}$ (A. Escala 2026), leads  to the simpler formula $\rm R/R_{\star} \approx \frac{1}{\sqrt{lnN}} \, M/M_{\star}$  valid for $t_{coll} =  t_{relax} $. We see that this dotted line passes thru  both intersections (between solid and dashed lines) on Fig 2b, as expected.

Fig 2b shows in red points the positions of LRDs for their upper limit sizes (i.e. same as in Figs 1 and 2a), in green   LRDs positions assuming that their real  effective radii corresponds to the 10\% of the   upper limits and in cyan, assuming that their real  effective radii corresponds to the 1\% of the measured  upper limits. In the case of the green points (10\% of the measured  upper limits), a fraction enters to the  `Forbidden Stellar Zone' ( $ t_{coll} <  t_{H0} < t_{relax}$), where it is expected that most of its  mass end-up forming a massive black hole, with a minor fraction of stars possibly escaping  from the system (i.e. becoming unbound in the formation process). The case of the cyan points (1\% of the measured  upper limits) it  is clearly more extreme, with most LRDs are in such situation of ending as a whole forming a MBH. In the latter case, a big population of MBHs of  $ \gtrsim  10^{10} \msun$ it  is predicted, which is currently unobserved in the Universe, suggesting  that the real radii should not be typically much less than 10\% of the observed  upper limits, or that LRDs  masses are considerably overestimated.  

A third (and less extreme) possibility  is explored in Fig 2c, assuming  that  it is the core radius of the LRDs the one equals to the 10\% (or 1\%) of their  measured (total) effective radii. Assuming a Plummer profile for the LRDs, the core radius can be defined where the projected mass density drops to half its central value,  which corresponds to  $r_c=a\sqrt{\sqrt2 -1}\approx$ 0.64$a$ (Dejonghe 1987, being  $a$  the Plummer radius), with an enclosed core mass $M(<r_c)= M_{tot} r_c^3/(r_c^2 + a^2)^{3/2}$ being equal to  15.7\%  of their total masses $M_{tot}$.  Our choice of a not so centrally concentrated profile (i.e. a Plummer instead of a more concentrated King model), is justified for LRDs since  $ t_{H0} < t_{relax}$  for theses systems and therefore,  their internal structure it is settled by mechanisms that operates on crossing times, such as  phase mixing and violent relaxation (Binney \& Tremaine 2008), instead of 2-body relaxation  where a steeper profile is expected. Fig 2c shows  in  green points  LRDs cores with radius  equals to the 10\%  of their  measured effective radii  and the cyan ones, equals to the 1\%  of their   effective radii.  For this case, from a fraction of LRDs cores enters to the `Forbidden Stellar Zone' (10\% core radius case), to most of them for the case where LRDs cores have radius  equals to the 1\%  of their  measured effective radii (in both cases it is assumed a core mass being equal to  15.7\%  of their total masses). Therefore, in this scenario the LRDs should be forming MBHs mostly in the range of $ \sim  3\times
10^7-3\times10^{10} \, \msun$ (if all the core ends up collapsing onto a MBHs, with lower masses if a fraction of stars scapes from the core in the process). In the case that most of the  core collapses onto a MBH,  they will  be over-massive respect to their host in a fraction similar to the one expected in the interpretation that LRDs already host a MBH ($\epsilon_{BH} \sim$ 0.1; F. Pacucci et al. 2023; K. Inayoshi \& K. Ichikawa 2024; E. Durodola et al. 2025).%, as in the case studied in Fig 2a.

It is important to note, that this new `core collapse' (possibly forming directly a MBH) differs from the standard one studied in stellar systems, since it is not driven by 2-body relaxation (because $t_{relax} >   t_{H0}$) and instead, directly by runaway stellar collisions. The quantitative difference with the 2-body relaxation  standard `core collapse',  will be that this    `core collapse'  happens without an  expansion of the envelope. For the same reason,  $t_{relax} >   t_{H0}$, it is unlikely that binary-single  interactions could halt the collapse, as well as other processes like  supernova feedback from massive stars segregated  to the core (not very effective against a collapsing system of point masses, like behaves stars on collapse scales) or the interaction with BHs that  segregate to the center (that would probably enhance the instability producing even more mergers with them). Therefore, it seems reasonable to conclude that a collapse of at least the core  appears unavoidable under the stellar-only interpretation for LRDs (Akins et al 2024).

In summary, in the three different possibilities analyzed here,  the formation  of a MBH seems  unavoidable, for all the discussed possible scenarios under the stellar-only interpretation for LRDs.  This is mainly due to the    extreme stellar  densities in LRDs inferred from JWST's observations, being almost unavoidable the onset of violent stellar collision that triggers a global instability of their cores (at the very least), probably forming MBHs directly in the supermassive range. Another recent work, F. Pacucci et al 2025, also predicts the formation of BHs in LRDs but in a considerably lower mass range ($\rm 10^{3-4} \msun$), using a variety of  approaches that, however, do not include %explore
 the role of the %(extremmelly short) 
collision timescale in their analysis. These approaches ranges from more appealing ones like N-body calculations  of dense stellar systems (for a very limited space parameter though: a single initial condition with evolution and result very similar to M.C. Vergara et al 2025), to not so promising ones, like computing the the expected LRD's restructure due to 2-body encounters in a system with relaxation time a factor 1,000 larger that the current age of the Universe (factor 10,000 for the Universe's age %Universe 
at the redshifts of LRDs;  F. Panucci et al 2025), being this also in agreement with our findings   that  typically $t_{H0} <<  t_{relax}$  for  LRDs. Our aim instead is to give a more global picture, based on the relevant timescales in the stability of LRDs  (i.e. systems with $t_{age} \sim \, t_{coll} < t_{relax}$),  in the full variety of properties displayed in their space parameter (for  the stellar-only interpretation; Akins et al 2024).

\section{Summary and Conclusions}

We study the stability and possible fates for LRDs, under the stellar-only interpretation of their observational features. We find that %some of
 these objects   typically  have $t_{age} \sim \, t_{coll} < t_{relax}$, therefore, in an unexplored  regime   observed for the first time since it was hypothesized   for a stellar system  six decades ago (Spitzer \& Saslaw 1966; Spitzer \& Stone 1967) and potentially, highly unstable to runaway collisions. We conclude  that, since LRDs are probably the densest stellar systems ever observed and  their cores, the  closest to be globally  unstable thru runaway collisions,  their nucleus are  the most favorable places to find a MBH in the process of formation (or recently formed). We reach to this conclusion after  analyzing three different scenarios in the stellar only interpretation, namely, (i) the numerical prediction of the densest stellar systems simulated so far,  (ii)  that LRDs real effective radii  corresponds to a fraction of the measured upper limits and (iii)  that the LRDs core radii corresponds to a fraction of the upper limit sizes, all with the same conclusion: the formation of a  VMS by catastrophic runaway stellar collisions in their  nuclear  region and subsequently, the possible formation of a MBH from the  direct collapse of such VMS.

It is important to note, that we reach  this conclusion through an  analysis of the relevant timescales and the comparison  of their observed properties with  recent numerical results of the densest stellar systems studied so far (M. C. Vergara et al 2023, 2024, 2025a, b), without the aid any fitting procedure or adjustment of a free parameter. Fitting procedures are restricted  to  the previously determined parameters  of LRDs properties (under the stellar-only interpretation) in Akins et al (2024), their observed  redshift range (relevant to give an upper limit for the age of the system; D. D. Kocevski et al. 2024)   and the black hole formation efficiency numerically determined in  M.C. Vergara et al (2025b) on stellar systems different from LRDs, but not on any part of the analysis  in this paper performed for LRDs  data.

Since the  alternative interpretations for LRDs have either a MBH already existing (R. Maiolino et al. 2024; J. E. Greene et al. 2024) or in the stages of final collapse  (L.  Zwick et al. 2025; J. Bellovary 2025) and considering that,  we showed (under the stellar only interpretation)  that LRDs  are  the most favourable known places to find a MBH in the process of formation% (VMS formation,  in the stages of final collapse, or just recently formed, depending on their exact place on the evolutionary stage described)
. This unifies the different interpretations for LRDs, into a context of ongoing MBH formation. Therefore, we conclude that under the different interpretations for LRDs, they  are candidates to  be the most appealing places to host the  massive seeds that will later produce   the observed population of high redshift quasars, besides the interpretation chosen for their observed properties.

% Therefore, in any of the two  the different interpretations for LRDs it is possible to  reach to the same conclusion: to  be the most appealing places to host the  massive seeds that led to the high redshift quasars., 

%also explaining  

%Depending o
Among   the different  interpretations, %will be if  
LRDs  %are mainly 
could be  on a stage of runaway stellar collisions  and early VMS  formation, already host a VMS in the stages of final growth and subsequent collapse, or with a MBH just recently formed, depending on their exact place on the described evolutionary stage. Since  LRDs typically appear in the Universe when $t_{H0} \sim$ 0.5 Gyr and disappear when $t_{H0} \sim$  1.5 Gyr, therefore having  maximum (individual) lifespan of  $\sim$ 1 Gyr (with more likely individual lifespans of few $10^8$yr or even less), the timescale for the instability that triggers global/core collapse is %instantaneous compared to such a short timescale (i.e the free fall time is many orders of magnitude shorter), being 
 probably comparable to their lifespans than  instantaneous  to them, therefore,  coexistence of the different stages within the whole LRDs population can also be expected. The lack of X-ray detections in most of those systems (T. T. Ananna et al. 2024; M. Yue et al. 2024; R. Maiolino et al. 2025), %only 
%that happens for the MBHs already formed, 
 supports that  most LRDs  to be in the earlier parts of the mentioned evolutionary path, being visible in X-rays only after the VMS collapses and a MBH is already formed, when the X-ray emission is  intense enough to escape from the surrounding  dust.

 The different evolutionary stages predicts different LISA gravitational wave observations (Amaro-Seoane et al. 2013), since  a gravitational-wave signal is expected from the final collapsing VMS at the moment of MBH formation, that will be detectable in the LISA band out to high redshift (Sun et al. 2017). Therefore,  gravitational wave observations could be discriminator for individual LRDs, among the different evolutionary stages. 
In addition, the extreme stellar densities needed for this instability could also be happening  in merging galactic systems (L. Mayer et al 2010, 2015), like in the almost face-on encounter of the recently discovered infinite galaxy, where it is suggested that a  MBH  is formed in-situ (P. Van Dokkum et al 2025a). Since the MBHs from the  original galaxies are also detected (P. Van Dokkum et al 2025b), this opens the additional possibility for  multiple LISA detections from  MBHs mergers (J.J. d'Etigny et al. 2024), from a single galactic system.

Finally, a relevant issue not studied at all  in this paper is the role of gas in the %formation and 
evolution of LRDs, particularly, in the enhancement of stellar collisions. Although the details of the exact evolution of the gaseous and stellar material are  unclear under realistic conditions, results from idealized calculations looks promising (T. Boekholt et al 2018, B. Reinoso et al 2023, 2025) and the dissipative nature of the gaseous component should certainly enhance instabilities. In particular,  the critical mass $\rm M_{crit}$ condition ($\rm t_{coll} \,(M_{crit}) = t_{H0}$) can be generalized  to include the presence  external potential from a gaseous component, which effectively reduces $\rm M_{crit}$ as it increases the velocity dispersion in the stellar system and therefore, increases the likelihood for collisions  (M.C. Vergara et al 2024). Most likely, the idealized case of extremely dense, purely gas-free LRDs discussed  in this paper rarely exists in the early universe, and most often, an unstable LRDs will (partially) collapse %as a whole 
during its formation before evaporating its gaseous envelope, as was already proposed  in  Escala (2021) for (proto-)NSCs at high z. The transitory nature of LRDs in a cosmological context, being a population observed during a cosmic period of only $\sim$1 Gyr in total (i.e. between z $\sim$ 4 and 8; D. D. Kocevski et al. 2024) and also, the extreme stellar densities inferred under the stellar-only interpretation (Guia et al. 2024), makes LRDs the most appealing places for hosting this (transitory and globally) unstable galactic nuclei.

%\newpage

We thank the anonymous reviewer for positive feedback. AE and SV acknowledges partial support from the Center of Excellence in Astrophysics and Associated Technologies  (FB210003). L.Z. acknowledges funding through the Alan S. Tetelman 1958 Fellowship for International Research in the Sciences. MCV acknowledges funding through ANID (Doctorado acuerdo bilateral DAAD/62210038) and DAAD (funding program number 57600326). MCV acknowledges the International Max Planck Research School for Astronomy and Cosmic Physics at the University of Heidelberg (IMPRS-HD). DRGS gratefully acknowledges support by the ANID BASAL project FB21003 and thanks for funding via the Alexander von Humboldt - Foundation, Bonn, Germany. ML acknowledges financial support from ANID/DOCTORADO BECAS CHILE 72240058.

%--------------------------------------------------------------------------------------------------------------

\end{document}